%% file: paper4.tex
\documentclass[useAMS,usenatbib]{mn2e}
\usepackage{graphicx}                                            
\usepackage{times}
\usepackage{color}

\usepackage{float}
\usepackage{rotating}
\usepackage{url}

\include{def_bibtex}
\include{definitions}

\title[Mass constraints to Sco X-1]{Mass constraints to Sco X-1 from Bowen flourescence and deep \\near-infrared spectroscopy}
\author[D. Mata S\'anchez et al.]{D. Mata S\'anchez$^{1,2}$, T. Mu\~noz-Darias$^{1,2,3}$, J. Casares$^{1,2}$, D. Steeghs$^{4}$, C. Ramos Almeida$^{1,2, \dagger}$ \newauthor and J. A. Acosta Pulido$^{1,2}$ \\
$^{1}$Instituto de Astrof\'isica de Canarias, 38205 La Laguna, Tenerife, Spain\\
$^{2}$Departamento de astrof\'isica, Univ. de La Laguna, E-38206 La Laguna, Tenerife, Spain\\
$^{3}$Department of Physics, Astrophysics, University of Oxford, Keble Road, Oxford, OX1 3RH, United Kingdom\\
$^{4}$Department of Physics, University of Warwick, Coventry CV4 7AL, UK\\
$^{\dagger}$ Marie Curie Fellow\\
}

\begin{document}
\maketitle

\begin{abstract}
More than 50 years after the dawn of X-ray astronomy, the dynamical parameters of the prototypical X-ray binary Sco X-1 are still unknown. We combine a Monte Carlo analysis, which includes all the previously known orbital parameters of the system, along with the K-correction to set dynamical constraints to the masses of the compact object ($M_1<1.73 $ M$_\odot$) and the companion star ($0.28\, M_\odot<M_2<0.70\, M_\odot$). For the case of a canonical neutron star mass of $M_1\sim 1.4 $ M$_\odot$, the orbital inclination is found to be lower than 40$^{\circ}$. We also present the best near-infrared spectrum of the source to date. There is no evidence of donor star features on it, but we are able to constrain the veiling factor as a function of the spectral type of the secondary star. The combination of both techniques restricts the spectral type of the donor to be later than K4 and luminosity class IV. It also constrains the contribution of the companion light to the infrared emission of Sco X-1 to be lower than 33\%. This implies that the accretion related luminosity of the system in the K band is larger than $\sim 4\times 10^{35}$~erg~s$^{-1}$.
  
\end{abstract}


\section{Introduction}
\label{intro}

Low-mass X-ray binaries (LMXBs) harbour a low-mass donor star which transfers matter onto a black hole or a neutron star (NS) via an accretion disc. They can be divided into two populations according to their long-term behaviour. \textit{Transient} systems spend most part of their lives in a faint, quiescent state, but show occasional outburst, where their X-ray luminosity increases above $\sim 10 \%$ of the Eddington luminosity (\ledd). There is also a population of $\sim$ 200 sources that are always X-ray active, displaying luminosities above $\rm L_X \simeq 10^{36} erg\, s^{-1}$ (but see \citealt*{Armaspadilla2013}). The so-called \textit{persistent} sources mostly harbour NS accretors, whilst the vast majority of black holes are found in transient systems. Transient LMXBs have provided a number of dynamical NS and black hole mass measurements thanks to the detection of the donor star during the quiescent phase (e.g. \citealt*{casares2014}).  This is not the case for the majority of the persistent population, where the companion spectrum is totally swamped by the reprocessed light from the accretion flow. Only for a few long orbital period systems a full orbital solution exists, since they harbour giant (i.e. bright), companion stars. 

Due to the blue spectrum of the accretion discs, and the typically late spectral types of the donor stars, near-infrared (NIR) observations offer a good opportunity to detect the companion star spectral features (e.g. \citealt{bandyopadhyay1997}) and obtain a full dynamical solution (see \citealt{steeghs2013} for a study purely based on NIR data).

Scorpius X-1 is the prototype LMXB and also the brightest persistent X-ray source in the sky, being the target of numerous studies since its discovery \citep{giacconi1962}. Although the spectral classification of the companion star is unknown, its relatively long orbital period ($P_{\rm{orb}}=18.9\, \rm h$, \citealt*{gottlieb1975}) suggests an evolved, late type star. \citet*{bradshaw1999} measured the trigonometric parallax of Sco X-1 using VLBA radio observations, and deduced a distance of $2.8\pm 0.3 \, \rm kpc$. Further observations allowed the detection of twin radio lobes, leading to an inclination value of $44^\circ\pm 6^\circ$. This value assumes that the radio-jet is perpendicular to the orbital plane \citep*{fomalont2001}. Even though the system does not show any of the NS identifying features (thermonuclear X-ray bursts or pulsations), a NS accretor is widely assumed based on its X-ray behaviour (see \citealt{vanderklis2006}; \citealt{Munoz-Darias2014} for X-ray emission from NS in LMXBs). 

\cite{steeghs2002} presented an optical spectroscopic technique for measuring system parameters in persistent LMXBs. It is based on the discovery of narrow emission lines within the Bowen blend, arising from the irradiated face of the donor star in Sco X-1 and powered by fluorescence. This claim is based on the narrowness, Doppler velocities and phase of the lines. Also, timming studies of the fluorescence emission have shown a time lag with the irradiating X-ray flux which is consistent with the light time between the two components of the binary \citep{Munoz-Darias2007}. 
Radial velocity studies of the Bowen lines (Bowen technique hereinafter) have been successfully applied to other dozen LMXBs \citep[see][]{cornelisse2008,mdariasth2009}, providing the first dynamical solutions to some classical NS systems \citep[e.g.][]{Casares2006, Cornelisse2007b} and the canonical black hole transient GX~339-4 \citep{Hynes2003,Munoz-Darias2008}.
The velocity inferred from these emission lines ($K_{\rm{em}}$) corresponds to that of the irradiated region of the companion; hence it only represents a lower limit to that of its centre of mass ($K_2$). In order to correct $K_{\rm{em}}$ from this effect numerical solutions should be applied, such as those computed by \citet*[][see Sec. \ref{kcorr}]{mdarias2005}.

In this paper we aim at constraining the masses of the components in Sco X-1 by combining two different techniques: (i) a K-correction along with Monte Carlo analysis accounting for constrains to the orbital parameters from previous studies (Section \ref{kcorr}) and (ii) NIR intermediate-resolution spectroscopy obtained around orbital phase zero, when the non-heated face of the donor star is oriented towards the Earth (Section \ref{irspec}). We note that Sco X-1, besides being the prototypical LMXB, is also an important object in the search for persistent gravitational waves, which also requires an accurate determination of its system parameters (e.g. \citealt{abadie2011}; \citealt{galloway2014}).

\section{The K-correction}
\label{kcorr}

The companion star of Sco X-1 has been solely detected through the Bowen technique described above. A \kem velocity was initially reported by \citet{steeghs2002}, and very recently refined by \citet[][see Table \ref{tableparam}]{galloway2014}. This velocity can be translated to that of the centre of mass of the donor by the so-called K-correction ($K_c$). \citet{mdarias2005} presented numerical solutions for $K_c$, which is parametrized by the following equation:
 
\begin{equation} \label{eq1}
K_c=\frac{K_{\rm{em}}}{K_2}  \cong N_0 +N_1 q +N_2 q^2 +N_3 q^3+N_4 q^4
\end{equation}
where $q=\frac{M_2}{M_1}$, is the mass ratio ($M_1$ and $M_2$ are the masses of the compact object and the donor star, respectively). The $N_i$ values are tabulated and depend on both the opening angle of the accretion disc that shades the companion ($\alpha$), and the orbital inclination ($i$). Therefore, $K_c$ is always lower than 1, and approaches unity as $\alpha$ increases and the regions of the companion with lower radial velocities become shaded. For the limit case in which the companion is fully shaded by the disc (i.e. $\alpha=\alpha_{M}$), $K_c$ is at maximum and can be approximated by the following expression.  

\begin{equation} \label{eq2}
 K_c = 1-f(q)\cdot (1+q) \qquad 
 f(q)\cong 1-0.213\, \left(\frac{q}{1+q}\right)^{\frac{2}{3}}
\end{equation}

\begin{figure}

\includegraphics[width=85mm]{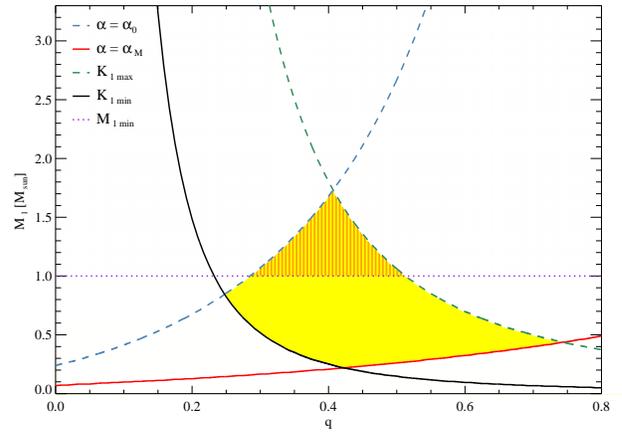}
\caption{K-correction for Sco X-1 obtained through a Monte Carlo simulation ($10^6$ events). We plot in solid and dashed, coloured lines the obtained limits to $M_1$ as a function of $q$. The upper limits correspond to $K_c(\alpha_0)$ (blue, dashed line) and $K_{1\rm{max}}$ (green, dashed line). The lower limits are defined by $K_c(\alpha_{M})$ (red, solid line) and $K_{1\rm{min}}$ (black, solid line). The yellow band represents the 90\% probability region delimited by the previous limits. The violet, dotted line corresponds to the imposed $M_1>1\, M_\odot$ constraint (see text), which restrict the allowed parameter space to the orange, stripped triangle.
}

\label{figMC}
\end{figure}

Therefore, for a given $q$ and $K_{\rm{em}}$, the true $K_2$ velocity is in the range:

\begin{equation} \label{eq3}
K_{\rm{em}}/ K_c(\alpha_{M})<K_2<K_{\rm{em}}/ K_c(\alpha_0)
\end{equation}
where $K_c(\alpha_0)$ represents the case in which the disc is not present and the inner face of the companion star is fully irradiated ($\alpha=0^\circ$). This gives the maximum K-correction and,  therefore, $K_c$ would be minimum.

\begin{figure*}
\includegraphics[width=158mm]{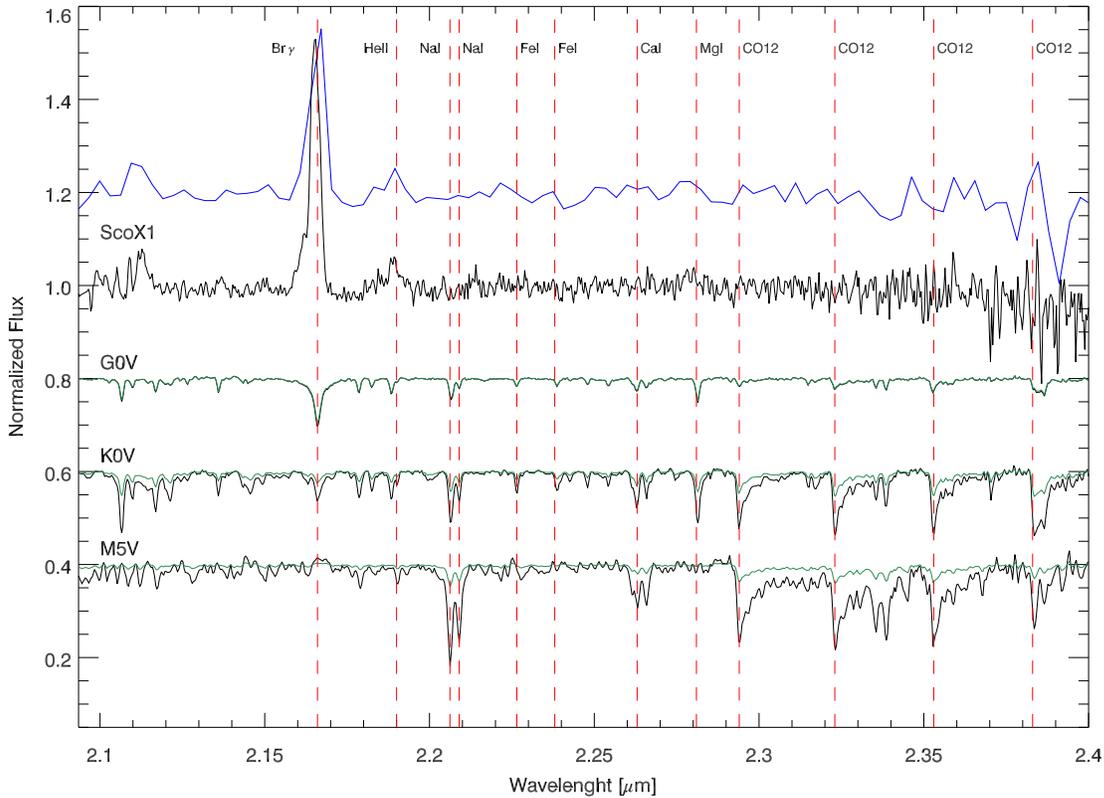}
\caption{Normalized NIR spectrum of Sco X-1 obtained by averaging the 7 nodding cycles (black solid line). For comparison we show (blue solid line; positive offset applied) the spectrum degraded to the resolution of the best NIR spectrum available so far \citep{bandyopadhyay1997}. We also include the spectra of three template stars (negative offsets applied) from IRTF Spectral Library as black and green solid lines. The latter have been veiled by the limit values reported in Table \ref{tableveil} for its corresponding spectral type. Typical NIR spectral features are identified by vertical, red, dashed lines. 
}
\label{figspec}
\end{figure*}
\subsection{Monte Carlo analysis}

The mass of the compact object ($M_1$)  can be inferred using the mass function formula, which is directly derived from Kepler's laws:
 \begin{equation} \label{eqmassfunc}
M_1=\frac{ P_{\rm{orb}}}{2 \pi G} \frac{K_2^3(1+q)^2}{\sin^3{i}}
\end{equation}
For the case of Sco X-1, some of the above system parameters are poorly constrained (e.g. orbital inclination) and for some others we just have extremal values (see also Section \ref{discussion}). This is the case of $K_1$ and $K_2$, the radial velocities of the compact object and the donor star, respectively. To deal with this, we performed a Monte Carlo analysis in order to obtain reliable limits to the compact object mass. We simulated $10^6$ random, normally-distributed values of $i$ and \kem, and lower and upper limits to the radial velocity of the compact object ($K_{1\rm{min}}$ and $K_{1\rm{max}}$, respectively) centred at the values reported in Table \ref{tableparam}. We apply both $K_c (\alpha_0)$ and $K_c(\alpha_{M})$ to each simulated \kem\ value. 
Finally, we solved Equation \ref{eqmassfunc} for $10^2$ steps in $q$ (from $0$ to $0.8$), using the previously defined distribution of $i$ values, as well as the orbital period reported in \citet{galloway2014}. We obtained four $M_1$ distributions: two lower limits corresponding to $K_2 (\alpha_M)$ and $K_{1\rm{min}}$, and two upper limits from $K_2 (\alpha_0)$ and $K_{1\rm{max}}$. These limits are represented by solid lines in Fig. \ref{figMC} and are obtained by adopting the value of $M_1$ above or below 90\% of the points (for upper and lower limits, respectively) in each distribution and for every sampled value of $q$.

\begin{table}
\centering
\caption{Observational measurements included in the Monte Carlo analysis.}

\begin{tabular}{c c c c }
\hline
$K_{\rm{em}}[\rm{km\, s^{-1}}]$ & $K_{1\rm{min}}[\rm{km\, s^{-1}}]$ &
$K_{1\rm{max}}[\rm{km\, s^{-1}}]$   & $i[\rm{deg}]$ 
\\
\hline

\hline
$74.9	\pm 0.5^{(1)}$ & $40\pm 5^{(2)}$ & $53\pm 1^{(2)}$ & $44\pm 6^{(3)}$   
\\

\hline
\end{tabular}
\begin{flushleft} 
REFERENCES: (1) \cite{galloway2014}; (2) \cite{steeghs2002}; (3) \cite{fomalont2001}; 

\end{flushleft}
\label{tableparam}
\end{table}

\subsection{The masses of the NS and the companion}

The intersections of the solid lines in Fig. \ref{figMC} provide absolute limits to $M_1$ and $q$. They yield extremal values to the system parameters with, at least\footnote{The confidence level of these results is actually slightly higher since two 90 \% limits (apart from that of the inclination) are simultaneously verified.}, 90\% confidence level:

\begin{center}
$0.22\, M_\odot<M_1<1.73\, M_\odot$ \\

$0.24<q<0.75$ \\

$0.05\, M_\odot<M_2<1.30\, M_\odot$\\
\end{center}

\noindent therefore, we obtain a robust upper limit of $M_1<1.73\, M_\odot$, consistent with a NS accretor. The lower limit is, on the other hand, not very restrictive. Adopting  instead $M_{1\rm{min}}=1 \, M_\odot$ as a conservative minimum value for the mass of a NS (see e.g. \citealt{kiziltan2013}), we further restrict the mass ratio to $0.28<q<0.51$, and hence $0.28\, M_\odot<M_2<0.70\, M_\odot$.

\section{Near Infrared spectroscopy}
\label{irspec}

We observed Sco X-1 on 2008 June 28, using the Long-slit Intermediate Resolution Infrared Spectrograph (LIRIS, \citealt{Acosta-Pulido2003}, \citealt{Manchado2004}) at the Cassegrain focus of the 4.2-m William Herschel Telescope (WHT), in La Palma (Spain). We used the K-band grism (R=2500, $\rm 3.5\, \AA/pix$ dispersion) combined with a 0.75'' slit. The use of grisms as dispersers usually does not permit to vary the spectral range. In our case, the employed slit (l0p75ext) consists of 2 slits which are offset along the dispersion axis by about 115 pixels in both senses. This allowed us to redshift the nominal spectral range to $\rm 2.09-2.45 \, \mu m$, which includes the CO bandheads. 

We performed seven nodding cycles of 3 positions each, with an exposure time of $900\, \rm{s}$  per cycle at airmasses in the range $\rm \sim 1.4 - 1.5$. We cover orbital phases, $\varphi = 0.01-0.13$ (using the ephemerides from \citealt{galloway2014}) and the total exposure time was $1.75\, \rm{h}$. For the data reduction process we used \textsc{IRAF}\footnote{\textsc{IRAF} is distributed by National Optical Astronomy Observatories, operated by the Association of Universities for Research in Astronomy, Inc., under contract with the National Science Foundation.} \textit{lirisdr} task package\footnote{The \textsc{IRAF} \textit{lirisdr} task package is supported by Jose Acosta.}, as well as a modified version of the package \textit{xtellcor$\_$general} developed by \citealt{vacca2003} to perform the telluric correction to the final spectra (see \citealt{Ramos2009,Ramos2013} for a more detailed description of the data reduction process).

This strategy yielded seven (nodding averaged) NIR spectra of Sco X-1. Absorption features from the donor star are not evident in any of them, which were combined to produce a higher signal-to-noise average spectrum. Points deviating more than $\rm 3\sigma$ above or below the continuum, probably caused by bad pixels or residual from the sky subtraction, were interpolated. The combined, normalized spectrum is presented in Fig. \ref{figspec}. To our knowledge this is the best NIR spectrum of Sco X-1 available to date (see \citealt{bandyopadhyay1997} for a previous example). No absorption lines from the companion star are detected. The most prominent feature in the spectrum is a strong $\rm Br\, \gamma$ emission line, whose full-width-at-half-maximum, $\rm{FWHM} =480\pm 30 \, \rm{km/s}$, is consistent with that of $\rm H \alpha$ ($\rm{FWHM} =440\pm 20 \, \rm{km/s}$). This latter value was obtained from previous observations performed by our team with the Intermediate Dispersion Spectrograph (IDS) at the Cassegrain focus of the 2.5-m Isaac Newton Telescope (INT), in La Palma (Spain).

\subsection{Veiling factor} 

Since no features from the companion are evident in our NIR spectrum, we investigated the amount of flux from the accretion flow required to veil the donor. This obviously depends on its spectral type. As a first step, we compared our spectrum with templates of G--M main sequence stars from the \textit{IRTF Spectral Library\footnote{\url{http://irtfweb.ifa.hawaii.edu/~spex}}}. In Fig. \ref{figspec}, we present normalized spectra of G0V, K0V and M5V stars, which cover the same spectral region as that of Sco X-1. For each spectral type, we define $F_{\rm{depth}}$ as the normalized flux of the deepest photospheric absorption line present in the NIR spectrum. These correspond to NaI ($\rm 2.206\, \mu m$), MgI ($\rm 2.282\,\mu m$) and $\rm { }^{12}CO$ ($\rm 2.295\, \mu m$), for M5V, G0V and K0V, respectively. Similarly, we define the veiling factor ($X$) as the fractional contribution of
the accretion related luminosity ($L_{\rm{acc}}^K$) to the total flux. The veiling necessary
to make these features shallower than the noise level ($3\sigma$) within the corresponding
spectral region of the Sco X-1 spectrum is a lower limit to this factor. Therefore,

 \begin{equation} \label{eq6}
X\ge 1- \frac{3\sigma}{F_{\rm{depth}}} \qquad X = \frac{L_{\rm{acc}}^K}{L_{\rm{acc}}^K+L_2^K}
\end{equation}
where $L_2^K$ is the luminosity of the donor. $X$ is always in the range $0 \leq X < 1$ and depends on the spectral type of the companion. Our lower limits to the fractional contribution of the accretion luminosity of Sco X-1 NIR flux ($X_{\rm min}$) are reported in the first column of Table \ref{tableveil}. It varies from 0.09 to 0.78 for the cases of G0V and M5V, respectively.

\section{Discussion}
\label{discussion}

The K-correction and Monte Carlo analysis (see Sec. \ref{kcorr}) impose an upper limit to the compact object mass of $M_1<1.73\, M_\odot$. This rules out a massive neutron star ($\sim 2\, M_\odot$) like e.g. that presented in  \citet{Demorest2010}. More accurate values of $\alpha$, $i$ and $K_1$ are necessary to improve this constraint. \citet*{deJong1996} proposed an average value of $\alpha=12^\circ$ for persistent LMXBs. Using this value, we can calculate $K_2$ by applying the same methodology, which constrains the mass of the compact object to $M_1<1.25\, M_\odot$. This value is lower than the canonical NS mass ($M_1\sim 1.4\, M_\odot$). However, we note that NS masses down to $M_1\sim 1.2\, M_\odot$ have been reported (e.g. \citealt{kiziltan2013}). If we fix the NS mass to the canonical value, we obtain $\alpha \leq 11^\circ$ and $K_1>47$ km s$^{-1}$. All the above constraints require an inclination as low as $i\sim36^{\circ}$ (we are quoting 90\% confidence limits).  

The upper limit to $K_1$ used in this work was obtained by applying the double Gaussian method to the \he{II} lines, while the lower limit comes from Doppler tomography analysis of \hb $\,$ \citep{steeghs2002}. Our analysis rules out this lower limit for a NS mass higher than $1\, M_\odot$ (see Fig.\ref{figMC}), and suggest the upper limit to be close to the real value. Interestingly, the latter stands true for another NS system analysed using the double gaussian technique, namely X1822-371 (Mu\~noz-Darias et al. in prep.; see also \citealt{Casares2003}), in which the $K_1$ velocity is accurately known \citep{Jonker2001}. Fixing $M_1=1.4\, M_\odot$ and assuming $K_1=K_{1\rm{max}}$, we obtain an absolute upper limit of $i\sim40^{\circ}$, which is only satisfied if $\alpha=0^\circ$ (i.e. applying $K_c(\alpha_0)$). Larger values of the opening angle or lower values of $K_1$ would result in even lower inclinations. We note that inclinations $\gtrsim 50^{\circ}$ were favoured to better explain time lags between reprocessed light (Bowen emission) and X-rays observed in Sco X-1. This value was found when the time-lags were compared with those predicted by numerical transfer functions with a maximum response. However, we note results consistent with those presented here are also obtained if a broader range of expected delays and/or more conservative solutions are considered (see \citealt{Munoz-Darias2007} for details).   

Finally, since Sco X-1 is not eclipsing, we can use our constraints to the mass ratio ($q$) and the Roche lobe radius (derived through the formulae in \citealt{paczynski1971}) to infer a maximum inclination purely based on this condition (i.e. absence of eclipses). This yields $i<70^\circ$. 

\subsection{The nature of the donor star}

Although our observations were performed at the most favourable orbital phase, we fail to detect any companion star feature in the NIR spectrum of Sco X-1 (see Sec. \ref{irspec}). Nevertheless, we obtain a lower limit to the veiling factor as a function of the spectral type of the secondary, and this can be used to discuss its nature. We initially found $0.05\, M_\odot<M_2<1.30\, M_\odot$, and assuming an accretor heavier than $1\, M_{\odot}$, we infer $0.28\, M_\odot<M_2<0.70\, M_\odot$. \citet*{faulkner1972} presented a direct relation between the mean density of the companion star and the orbital period in the Roche lobe filling binaries. For a main sequence companion in a 18.9 h orbital period, this result in a spectral type A0 \citep{cox2000}. However, the features of an A0V donor would dominate the observed spectrum. Similarly, its mass ($M_2= 2.9\, M_\odot$) is ruled out by our results. This implies that the donor star is not a regular dwarf star but a somewhat more evolved object. Our upper limit ($M_2<0.70  M_\odot$) contraints the spectral type to be later than K4, since this value corresponds to a main sequence donor. Using the same procedure as in Section \ref{irspec} for a K4 donor we obtain a lower limit to the veiling factor of $X>0.66$. 

We can independently test the allowed spectral types for the companion star by solely using the veiling factor constraints. In a first step, we use the absolute K-magnitude value of Sco X-1 $M_K^{\rm{tot}}=-1.14$ (see \citealt{wachter2005} and references therein) to set a minimum value of the absolute K-magnitude of the donor star as a function of the spectral type (see Table \ref{tableveil}). Then, we combine the values tabulated for $M_V$ and $(V-K)$ (see \citealt{cox2000} and \citealt{koornneef1983}, respectively) to infer absolute K-magnitudes for different spectral types and luminosity classes. We conclude that all the inspected spectral types would be compatible with the veiling constraints if the companion were in the main sequence ($M_{K}^{\rm{V}}$). However, doing the same for giant stars ($M_{K}^{\rm{III}}$) we are able to rule out luminosity class III (and earlier) for the donor. On the other hand, assuming that the companion star radius is at the Roche lobe \citep{paczynski1971}, we obtain subgiant K-magnitudes ($M_{K}^{\rm{IV}}$) compatible with our veiling constraints (see Table \ref{tableveil}).

\subsection{Accretion luminosity in the NIR}

The above constraint to the veiling factor ($X>0.66$) implies that the accretion luminosity ($L_{\rm{acc}}^K$) accounts for more than 2/3 of the observed K band flux. If we consider the K band luminosity of Sco X-1 reported by \cite{wachter2005}, and propagate the errors on the interstellar extinction and the distance, we obtain $8\pm 2\times 10^{35}$ erg s$^{-1}$. As noted by these authors, the error is not dominated by uncertainties in the measurements but reflects real variability ($\sim 25 \%$) in the K band. According to the MAXI all sky monitor \citep{Matsuoka2009}, the day of our observations Sco X-1 flux was within 10\% of its long-term average. Since data were obtained when the donor contribution is at maximum, we can use the limit $X>0.66$ to constrain the accretion luminosity of Sco X-1 in the NIR, which is therefore in range $4\times 10^{35}\,$ erg s$^{-1} < L_{\rm{acc}}^K< 10^{36}\, $ erg s$^{-1}$.

\begin{table}
\centering
\caption{Minimum veiling factor ($X_{\rm{min}}$) and implied minimum K-magnitude ($M_{K\, \rm{min}}$) for the donor star. $M_{K}^{\rm{V}}$, $M_{K}^{\rm{IV}}$ and $M_{K}^{\rm{III}}$ refer K-magnitudes of dwarf-to-giant standard stars.} 
\begin{tabular}{c c c c c c }
\hline
Sp. Type & $X_{\rm{min}}$ & $M_{K\, \rm{min}}$   & $M_{K}^{\rm{V}}$ & $M_{K}^{\rm{IV}}$ & $M_{K}^{\rm{III}}$ 
\\
\hline
\hline
G0 & $0.09$ & $-1.04$ & $3.18^{(1,2)}$ & $3.05^{(1,2,3)}$ & $-1.18^{(1,2)}$ \\
K0 & $0.61$ & $-0.12$ & $4.13^{(1,2)}$ & $3.89^{(1,2,3)}$ & $-1.65^{(1,2)}$\\  
M5 & $0.78$ & $ 0.51$ & $7.80^{(1,2)}$ & $7.06^{(1,2,3)}$ & $-6.50^{(1,2)}$
\\

\hline
\end{tabular}
\begin{flushleft} 
REFERENCES: (1) \cite{cox2000}; (2) \cite{koornneef1983}; (3) \cite{paczynski1971}

\end{flushleft}
\label{tableveil}
\end{table}  

\section{Conclusions}
\label{conclusion}

We performed a Monte Carlo analysis along with the K-correction to constrain the dynamical parameters of the prototypical X-ray binary Scorpious X-1. We obtain the following constraints, at 90\% confidence level:

\begin{center}
$M_1<1.73\, M_\odot$ \\
$0.28<q<0.51$ \\
$0.28\, M_\odot<M_2<0.70\, M_\odot$\\
\end{center}
Thus, the presence of a massive neutron star ($\sim 2\, M_\odot$) in this system is ruled out. If we consider standard values for LMXBs, a possible set of parameters would be a canonical NS of $M_1\sim 1.4\, M_\odot$, an orbital inclination of $i\sim 36^{\circ}$ and a disc opening angle of $\sim 11 ^{\circ}$. Assuming $M_1 = 1.4\, M_\odot$ necessarily implies $i \lesssim 40^{\circ}$. Higher values of the inclination would only be possible if the neutron star is lighter than the canonical value or any of the measurements used in this work is not correct. 

We also presented the best NIR spectrum of the source to date. Despite the non detection of donor star features, our deep observations constrain the spectral type of the donor to be later than K4 and luminosity class IV. We obtain a lower limit to the veiling factor in the NIR of $X>0.66$, implying that the accretion related luminosity in the K band is larger than a few times $10^{35} $ erg s$^{-1}$.

\section{Acknowledgements}
\label{acknowledgements}

DMS acknowledges Fundaci\'on La Caixa for the financial support received in the form of a PHD contract. TMD acknowledges funding via an EU Marie Curie Intra-European Fellowships under contract numbers 2011-301355. CRA is supported by a Marie Curie Intra European Fellowship within the 7th European Community Framework programme (PIEF-GA-2012-327934). The William Herschel Telescope is operated on the island of La Palma by the Isaac Newton Group in the Spanish Observatorio del Roque de los Muchachos of the Instituto de Astrof\'isica de Canarias. DMS and JC acknowledge the hospitality of the Department of Astrophysics of the Oxford University (UK) where part of this work was carried out. This paper is supported by the Spanish Ministerio de Econom\'\i{}a y Competitividad (MINECO) under grants AYA2010-18080 and SEV-2011-0187.

\bibliographystyle{mn2e} 
\bibliography{MiBiblio4.bib} 

\label{lastpage}
\end{document}

%% file: definitions.tex
\newcommand{\he}[1] {He\,{\sc #1}}

\newcommand{\hb}{H$\beta$}

\def\lesssim{\mathrel{\hbox{\rlap{\hbox{\lower4pt\hbox{$\sim$}}}\hbox{$<$}}}}
\let\la=\lesssim
\def\gtrsim{\mathrel{\hbox{\rlap{\hbox{\lower4pt\hbox{$\sim$}}}\hbox{$>$}}}}
\let\ga=\gtrsim

\def\kem{$K_\mathrm{em}$}

\def\ledd{$L_\mathrm{Edd}$}

